\documentstyle[12pt,epsf]{article}
\newcommand{\RR}{\mbox{${\rm \:  R\!\!\!\! I
\;\;}$}} 

\newcommand{\qed}{\hfill $\Box$ \vskip 2ex}
\newcommand{\be}[1]{\begin{equation}\label{#1}}
\newcommand{\ee}{\end{equation}}
\newcommand{\vs}{\vspace{0.25cm}}

\begin{document}

\begin{center}

{\Huge Control of One and Two Homonuclear Spins}

\end{center}

\bigskip

\begin{center}

{Domenico D'Alessandro \\ 
Department of Mathematics\\
Iowa State University, \\
Ames, IA 50011,  USA\\
Tel. (+1) 515 294 8130\\
email: daless@iastate.edu}

\end{center}

\vspace{0.5cm}

\begin{abstract}

We consider the problem of steering 
control for the systems  of one spin
$\frac{1}{2}$ particle and two 
interacting homonuclear spin $\frac{1}{2}$ particles 
in an electro-magnetic field. The describing 
models are bilinear systems whose
state varies on the Lie group of special 
unitary matrices of dimensions 
two and four, respectively. By performing 
decompositions of Lie groups,
taking into account the describing 
equations at hand, 
we derive control laws to steer the state 
of the system to any desired final configuration. 
Explicit formulas are 
given for the parameters involved in the control
algorithms. Moreover, the proposed 
algorithms allow for arbitrary bounds
on the magnitude of the controls and for some flexibility in the
specification of  the final 
time which must be greater than a given value but otherwise arbitrary. 

\end{abstract}

\vs 

\vs


\vs

\vs

\section{Introduction}
In recent years, there has been a great 
amount of interest in the study
of control of quantum mechanical systems 
(see e.g. \cite{BS}, \cite{conmoh}, 
\cite{Potzscr}, \cite{WRD}). This has been motivated by recent advances in the
area of nuclear magnetic resonance and laser spectroscopy which have rendered 
possible the introduction of  active control at the atomic
level. A major motivation to study control of quantum mechanical system
is given by quantum computation \cite{diviscience}. The information in a
quantum computer is encoded in the state of a quantum system which has
to be manipulated in order to initialize the computer, perform logic
operations and measure the result of the computation. Existing
techniques for the manipulation of the state of quantum systems only
allow to perform very simple operations.  On the other hand, 
the introduction of a control theoretic point of view promises to greatly increase the
number of operations that can be implemented with a quantum system as
well as their accuracy. 

\vs

The state of a general multilevel quantum 
system is described by a vector $|\psi(t)>$, 
in a finite dimensional Hilbert space. At every 
time $t$, we have  
\be{stato}
|\psi(t)>=X(t) |\psi(0)>.
\ee 
The operator  $X(t)$ is the evolution operator 
solution of the Schr\"odinger equation 
\be{scro}
i \hbar \dot X=H X, 
\ee    
with initial condition equal to the identity, and $H$ is the Hamiltonian 
operator. In many experimental situations,  such as nuclear 
magnetic resonance spectroscopy,  
the Hamiltonian $H$ 
has the form  
$H:=H_0+\sum_{i=1}^m H_i u_i(t)$ where 
$u_i(t)$ are externally applied 
electro-magnetic 
 fields which play the role 
of controls. This simplified 
model assume no
interaction with the environment 
other than  through the 
controls. In this  situation,  equation 
(\ref{scro}) can be written  as   
\be{bassys}
\dot X(t)=AX(t)+
\sum_{i=1}^mB_iX(t)u_i(t), 
\qquad X(0)=I_{n \times n}.  
\ee
In (\ref{bassys}),   
$A$, $B_1,...,B_m$,  
are matrices in $su(n)$, 
where $n$ is the number of 
levels of the system under consideration. 
The solution of
(\ref{bassys}),   
$X(t)$,  varies on the Lie group 
of special unitary matrices of dimension $n$,  $SU(n)$. In this paper,  
we  are interested 
in the control  of $X$ in (\ref{scro}). This 
is a  way to obtain 
the control   of the state $|\psi>$ 
in (\ref{stato}) and also to control 
the {\it operation} in (\ref{stato}) to 
be  performed on 
the state of the quantum system. 

\vs 

Decompositions of Lie groups, and in particular 
of the Lie group $SU(n)$, have been 
recently used to prescribe control 
laws and to study the controllability properties 
for multilevel quantum systems described 
by equations such as (\ref{bassys}) (see e.g. \cite{KBG}, 
\cite{KG},  \cite{Rama1},
 \cite{Rama2}, \cite{Reck}). 
To illustrate the basic idea, 
let us consider the simplest case of a 
two level quantum system described by  
\be{modelsu2}
\dot X= AX+BXu. 
\ee 
The matrices $A$ and $B$ are 
in the Lie algebra $su(2)$ 
and the initial condition is
assumed to be the identity. 
The matrices  $A$ and $B$ are linearly 
independent and therefore they generate the whole Lie
algebra $su(2)$, since $su(2)$ has no two dimensional subalgebras. 
\footnote{If $A$ and $B$ are not linearly independent, 
the solution of (\ref{modelsu2}),  with initial 
condition equal to the identity,  can be written as 
\be{sol}
X(t)=e^{At+B\int_0^t u(\tau) d\tau}, 
\ee        
and, to obtain a desired final 
configuration,  we can just select a control 
function $u$ to make ${At+B\int_0^t u(\tau) 
d\tau}$ equal to one 
of the logarithms of 
the desired target $X_f$, 
if $X_f$ is in the 
Lie subgroup of $SU(2)$, 
 described by $\{ X \in SU(2)| 
X=e^{B s}, s \in \RR \}$.}  
Define\footnote{The 
inner product $<\cdot, \cdot>$
is defined  as   $<A,B>:=Tr(AB^*)$ and it is equal to the Killing form
on $su(2)$ (see e.g. \cite{Helgason}).}  
\be{key}
k:=\sqrt{\frac{<A,A>}{<B,B>}}. 
\ee
The quantity $k$ is a measure of the 
`control authority' of the given system. The matrices $A+kB$ and $A-kB$
are orthogonal and therefore, if $T_1$ is the unitary matrix which
diagonalizes $A+kB$, we have 
\be{diagonalizzz}
T_1(A+kB)T_1^*=-i \lambda S_z,  
\ee
and 
\be{diagonalizzz2}
T_1(A-kB)T_1^*=-i(aS_y+bS_x),   
\ee
for some parameters $\lambda >0$, $a,b$ not both zero. $S_x$, $S_y$ and
$S_z$ are the Pauli matrices  
\be{Paulix}
S_x:=\frac{1}{2}\pmatrix{0 & 1 \cr 1 & 0},
\ee  
\be{Pauliy}
S_y:=\frac{1}{2}\pmatrix{0 & -i \cr i & 0},
\ee
\be{Pauliz}
S_z:=\frac{1}{2}\pmatrix{1 & 0 \cr 0 & -1}. 
\ee
The matrices 
$-iS_x$, $-iS_y$, and $-iS_z$ form an orthogonal  
basis in $su(2)$ and we have  the
basic commutation relations 
\be{commuPauli}
[-iS_x,-iS_y]:=-iS_z,  \qquad 
[-iS_y,-iS_z]:=-iS_x,  \qquad [-i S_z,-i S_x]:=-i S_y. 
\ee
Define the unitary matrix 
\be{T2}
T_2:=\pmatrix{1 & 0 \cr 0 & \frac{a+ib}{\sqrt{a^2+b^2}}}. 
\ee
We have 
\be{CoC1}
T_2T_1(A+kB)T_1^*T_2^*=-i\lambda S_z, 
\ee
\be{CoC2}
T_2T_1(A-kB)T_1^*T_2^*=-i\sqrt{a^2+b^2}S_y. 
\ee
In geometric terms, the transformation 
$T_1$ is a rotation which ensures
that the $z$ axis is aligned along 
the direction specified by $A+kB$ while
$T_2$ is a rotation about the $z$ 
axis to align the direction of rotation specified by 
$-i(aS_y+bS_x)$ with the $y$ direction. The change of coordinates $T_2T_1$
shows that it is always possible to assume that $A+kB$ and $A-kB$ are
proportional to the Pauli matrices $S_z$ and $S_y$, respectively.   

\vs

Now assume we want to find a control input steering to a final
configuration $X_f$ (after the change of coordinates). We can first
express $X_f$ using the Euler parametrization of matrices in $SU(2)$
(see e.g. \cite{sakurai}) 
\be{EULERO}
X_f:=e^{-i S_z \alpha} e^{-i S_y \beta} e^{-i S_z \gamma}, 
\ee
with $\alpha, \gamma \in [0,4\pi]$, $\beta \in [0,\pi]$. Once 
the parameters in
(\ref{EULERO}) are known, it is immediate to find a control function
steering the state of the system to the value $X_f$. We have the
following result

\vs

\noindent{\bf Theorem 1} {\it The control piecewise constant 
and equal to $k$ in the  interval $[0,\frac{\gamma}{\lambda}]$, 
$-k$ in the interval $( \frac{\gamma}{\lambda}, \frac{\gamma}{\lambda}+ 
\frac{\beta}{\sqrt{a^2+b^2}}]$ and equal to $k$ in the interval 
$( \frac{\gamma}{\lambda}+ 
\frac{\beta}{\sqrt{a^2+b^2}}, \frac{\gamma}{\lambda}+ 
\frac{\beta}{\sqrt{a^2+b^2}}+ \frac{\alpha}{\lambda}]$, steers the state
of the system (\ref{bassys}) (in the new coordinates) to $X_f$ in 
(\ref{EULERO}).} 

\vs
Results similar in spirit to Theorem 1 have appeared in  
\cite{Rama1} \cite{Rama2}. In particular the paper \cite{Rama2} contains
a factorization of the Lie group $SU(2)$ which allows control with
arbitrarily bounded power. The authors deal
with the case where $A$ and $B$ are orthogonal and then generalize to a
large class of system on $SU(2)$. 

\vs

Consideration of a bounded 
amplitude control is quite natural in
applications like nuclear 
magnetic resonance where the term $A$ in 
(\ref{modelsu2}) corresponds to a large 
 constant magnetic field (so that the value of $k$ in 
(\ref{modelsu2}) is typically large) while the time 
varying transverse magnetic 
field $u$ has small amplitude. 
In the next section, using a Lie 
group decomposition generalizing Euler 
decomposition (\ref{EULERO}) we 
will give a control algorithm for any 
system on $SU(2)$ of the form (\ref{modelsu2}) which allows for 
arbitrary bounds on the control magnitude. Sections 3 through 5 are
devoted to the more complicated case of two interacting spin
$\frac{1}{2}$ particles. Emphasis is given to the {\it homonuclear} 
case, which for us means that the two particles have the same
gyromagnetic ratios. We derive, in Section 3 the model of two interacting spin
$\frac{1}{2}$ particles and prove some properties of the Lie algebra
$su(n)$ which are relevant for the controllability of this system. The
controllability of the system of two homonuclear spin $\frac{1}{2}$
particles is dealt with in detail in Section $4$, where we give explicit
expressions for the reachable sets at every time. This
treatment gives information about what states can be reached at any
given time. In Section 5, we present an algorithm to steer exactly the state of
this system  to any prescribed final condition with arbitrary bounds on
the control amplitude. Conclusions are presented in Section 6.  


\section{Control of two Level quantum 
Systems with Arbitrarily Bounded Control}

Consider once again system 
(\ref{modelsu2}) that we rewrite here 
\be{modelsu23}
\dot X=AX+BXu, 
\ee 
with $A$ and $B$ in $su(2)$ 
and linearly independent, and 
$X(0)=I_{2 \times 2}$. Assume that the 
amplitude of the control $u$ is
bounded by $|u| \leq M$. We will 
consider a piecewise constant control
which attains only the values $\pm M$ 
(bang-bang). 
In order to specify
the switching times for the bang-bang control, we will consider 
a decomposition of the Lie group
$SU(2)$ in terms of  the matrices 
\be{Z1}
Z_1:=A+MB, 
\ee   
\be{Z2}
Z_2:=A-MB. 
\ee

\vs
In the following we assume that $Z_{1,2}$ are linearly independent
matrices in $su(2)$ and we consider the following parameters: 

\vs 

\begin{itemize}

\item $\lambda_{1,2} >0$ 
are defined as the magnitudes 
of the purely imaginary eigenvalues of 
$Z_{1,2}$. The one parameter subgroups
associated to $Z_{1,2}$ are periodic 
with period 
$\frac{2 \pi}{\lambda_{1,2}}$ and $\lambda_{1,2}$ is a measure of the
speed at which one moves on the one parameter subgroup corresponding to 
$Z_1$ and $Z_2$, respectively.

\item The parameter $\psi$ is the cosine of the
angle between $Z_1$ and $Z_2$, namely we define 
\be{psi}
\psi:= \frac{<Z_1,Z_2>}{<Z_1,Z_1>^{\frac{1}{2}}
<Z_2,Z_2>^{\frac{1}{2}}}. 
\ee 
The parameter $\psi$  is not changed by a change of coordinates
(rotations of the reference frame) nor by a
scaling of $\lambda_{1}$ and/or $\lambda_2$. If $\psi=0$ then $Z_1$ and
$Z_2$ are orthogonal. If and only if $\psi=1$,  $Z_1$ and $Z_2$ are
proportional to each other which we have excluded so that in general we
have $0 \leq |\psi| < 1$.  

\end{itemize}

\vs 

Let $T_1$ be the unitary matrix that diagonalizes $Z_1$ namely 
\be{diagZ1}
T_1Z_1T_1^*=-i2\lambda_1S_z, 
\ee
and define
\be{altro}
T_1Z_2T_1^*:=-i(aS_y+bS_x+cS_z). 
\ee
With $T_2$ given by 
\be{T2bis}
T_2:=\pmatrix{1 & 0 \cr 0 & \frac{a+ib}{\sqrt{a^2+b^2}}}, 
\ee
we have 
\be{new}
T_2T_1 Z_1 T_1^* T_2^*=-i2\lambda_1S_z, 
\ee
\be{ne11}
T_2T_1 Z_2 T_1^* T_2^*=-icS_z-i\sqrt{a^2+b^2}S_y.  
\ee
Therefore there exists a unitary matrix $W:=T_2 T_1$ such that 
\be{camcoor}
WZ_1W^*=-i2\lambda_1S_z, \quad WZ_2W^*=-icS_z-i\sqrt{a^2+b^2}S_y:=-iD.  
\ee

\vs

The parameters $\lambda_1$, $\lambda_2$, $\psi$ and the matrix 
 $W$ will all play a role in determining the generalized Euler 
parameters in the factorization of elements of $SU(2)$ 
in terms of $Z_1$ and $Z_2$ described in
the following theorem. 

\vs

\noindent{\bf Theorem 2.}{\it Consider 
an arbitrary (target) 
matrix in $SU(2)$, $X_f$ and 
let $\alpha \in [0, 4\pi]$,  
$\beta \in [0,\pi]$ and 
$\gamma \in [0,4\pi]$ be the Euler parameters of
the matrix $WX_fW^*$, namely 
\be{neueul}
WX_fW^*= e^{-i S_z \alpha} e^{-i S_y \beta} e^{-i S_z
\gamma}. 
\ee 
Choose a positive integer $m$ such  that
\be{admissibility}
cos^2(\frac{\beta}{2m}) \geq \psi^2. \ee 
Then $X_f$ has the following factorization: 
\be{FactXf}
X_f=e^{Z_1 \frac{\alpha}{2 \lambda_1}} 
(e^{Z_1t_1} e^{Z_2 t_2} e^{Z_1 t_3})^m 
e^{Z_1\frac{\gamma}{2 \lambda_1}}, 
\ee
with 
\be{t2}
t_2:=\frac{1}{\lambda_2} cos^{-1} \sqrt{\frac{1}{1-\psi^2} 
(cos^2(\frac{\beta}{2m})-\psi^2)}
\ee
and setting
\be{phi}
\phi:=tan^{-1}(-\psi tan (\lambda_2 t_2)), 
\ee
or 
\be{phibis}
\phi:=- sign(\psi) \frac{\pi}{2},
\ee
if $\lambda_2t_2=\frac{\pi}{2}$,
\be{t1}
t_1=t_3=\frac{\phi}{2 \lambda_1}, 
\ee
if $\phi \geq 0$, 
and  
\be{t3}
t_1=t_3=\frac{2 \pi +\phi}{2 \lambda_1},  
\ee
if $\phi<0$.}
\vs

The proof of the above decomposition is presented in \cite{PPR}. 
The  described decomposition of the Lie group $SU(2)$ and calculation of the parameters can be
easily extended to the Lie group $SO(3)$. The relevant calculations for the latter case 
are presented in Appendix $B$. 
\vs 

F. Lowenthal \cite{Lowe} first showed that the Lie group $SU(2)$ is
uniformly generated \cite{CSL1} \cite{CSL2} 
by any two linearly independent matrices in $Z_1$ $Z_2$.  This  means that
every element $X_f$ of $SU(2)$ can be written as 
the finite product  of alternate elements of the one
dimensional subgroups corresponding to $Z_1$ and $Z_2$,  namely 
\be{sfg}
X_f=e^{Z_1 t_1}e^{Z_2 t_2}e^{Z_1 t_3}e^{Z_2 t_4}\cdot \cdot \cdot e^{Z_1 t_s}, 
\ee  
for some parameters $t_1,t_2,...,t_s >0$ and that,   
although $s$ depends on
$X_f$, it is uniformly bounded over $SU(2)$. The contribution of Theorem 2
above is that we provide explicit formulas for the parameters $t_j$. 
The maximum number of factors `$s$' (maximum over all of $SU(2)$) is
called {\it order of generation} and it is the minimum number of factors
 needed to express all of the elements of $SU(2)$ as in 
(\ref{sfg}). The order of generation  depends on the angle $\psi$
defined as in (\ref{psi}),  
 between $Z_1$ and $Z_2$. In particular,  it
is minimal and equal to three if $Z_1$ 
and $Z_2$ are orthogonal.  
F. Lowenthal in \cite{Lowe} has derived a formula which relates 
$\psi$ to  the order of generation  $s$ 
in (\ref{sfg}). In particular,  $s=3$, 
if 
$\psi=0$ and $s=f+2$ if 
\be{Lwenthalform}
cos(\frac{\pi}{f}) < 
|\psi| \leq cos(\frac{\pi}{f+1}), 
\ee
with $f \geq 2$. 

\vs 

It is interesting to compare the number of factors for
the factorization described in Theorem 2 and the minimum required
according to Lowenthal formula (\ref{Lwenthalform}). The number of
factor required in (\ref{FactXf}) is $2m+1$ where $m$ is the smallest
positive integer satisfying (\ref{admissibility}). The worst case is
when $\beta=\pi$, which shows that the minimum number of factors needed
to express all the matrices $X_f$ in $SU(2)$ in the form (\ref{FactXf})
is $2m+1$ where $m$ is the smallest integer satisfying  
\be{admiss2}
|\psi| \leq cos(\frac{\pi}{2m}).  
\ee   
A comparison of (\ref{admiss2}) with (\ref{Lwenthalform}) shows
(identifying $f+1=2m$) that the
number of factors needed in our algorithm is exactly the one given by
F. Lowenthal, namely the smallest possible, in the case when $f$ in 
(\ref{Lwenthalform}) is odd and larger by just one if $f$ is even. In fact,
the derivation of Theorem 2 was in the spirit of  
F. Lowenthal proof of his uniform generation result in \cite{Lowe} 
\cite{Lowe2}. F. Lowenthal uses stereographic projections, 
translates the problem to the induced 
subgroup of the Moebius group and uses some of his previous 
results  \cite{Moebius}. Since we are  
interested in the determination
of the parameters involved 
in the factorization (\ref{FactXf}), we worked directly on the Lie group
 $SU(2)$ and derived the factorization using explicit expressions of the
matrices involved. 

\vs

Given the decomposition in Theorem 2, it is now immediate to find a
control algorithm to steer the state of system (\ref{modelsu23}). We
have the following: 

\vs

\noindent{\bf Theorem 3} {\it Consider the piecewise constant control
equal to $M$ for a time $\frac{\gamma}{2\lambda_1}$, and then equal to
$M$, $-M$, $M$ for times $t_3$, $t_2$ and $t_1$, respectively, $m$
times, and then equal to $M$ for an interval of time of length
$\frac{\alpha}{2 \lambda_1}$. This control steers the state $X$ of 
(\ref{modelsu23}) from the identity to $X_f$ in (\ref{FactXf}).} 

\vs

Notice that Theorem 3 is a direct generalization 
of Theorem 2 that can be obtained as a special
case if $\psi=0$, $m$ is chosen equal to 
 $1$ and one sets $2 \lambda_1  = \lambda$, 
$2 \lambda_2 = \sqrt{a^2 +b^2}$. Notice also that , although the
algorithm allows to reach a given state $X_f$ at a given time, say
$T_f$, one can consider the actual final time as arbitrary as long as it
is greater than a given value. This will be very important in the
algorithms of Section 5 where we use this flexibility to obtain a given
state at a given time in a suitable rotating frame. Let us illustrate
this point by assuming (w.l.g.) $Z_1=-iS_z$, with $S_z$ in
(\ref{Pauliz}). Then we have 
\be{hhh76}
e^{-iS_y \pi} Z_1 e^{iS_y \pi}=-Z_1. 
\ee
If we let the system evolve as  $e^{Z_1t}$, for time $\bar t$, then steer to $e^{iS_y \pi}$ in
time $T_1$, then let the system evolve   as $e^{Z_1 t}$ for time $\bar
t$ and then steer to $e^{-iS_y \pi}$ in time $T_2$, 
we obtain from (\ref{hhh76}) the identity matrix 
in time $2 \bar t + T_1 +T_2$.  If we
follow this procedure after having driven the state of the system to
$X_f$ in time $T_f$, we steer to $X_f$ in time $T_f+T_1+T_2+2 \bar t$,
with $\bar t$ arbitrary. More examples of the use of this procedure will
be given in Section 5.

\vs    

We conclude this section by discussing how the number 
of factors (switches) needed in the described
algorithm, which is, as we discussed above,  
essentially the minimum number needed, depends on
the amplitude of the control. Notice, from formula 
(\ref{admissibility}), that the number of factors $2m+1$ 
increases with the value of $|\psi|$.  
By substituting $Z_1=A+MB$, $Z_2=A-MB$ in $\psi$ 
in (\ref{psi}),   after some
elementary manipulations, we obtain
\be{Lowenthalfunction}
|\psi(M)|:=\frac{|k^2-M^2|}{\sqrt{(k^2+M^2)^2-4M^2(\frac{<A,B>}{<B,B>})^2}}. 
\ee
The parameter $k$ is defined in (\ref{key}). 
 A study of the function $|\psi|$ in (\ref{Lowenthalfunction}) shows that 
$M=k$ achieves the minimum value of number of switches. The function 
is decreasing in the interval $[0,k)$ and increasing in $[k,+\infty)$. 
The number of switches and the number of factors in (\ref{sfg}) tends to
infinity as $M$ goes to zero as well as $M$ goes to $+\infty$. 
Therefore,  in order 
to minimize the number of switches, if the control $u$ is bounded 
by $|u| \leq M$ and
$M>k$, it is convenient  to use $u =\pm k$ rather than $u=\pm M$. In
other terms, there is a natural value for the control, given by $k$ in
(\ref{key}) which is the best to use even though we are allowed higher
amplitude controls, as far as the number of switches is concerned.

\section{Model of two interacting spin $\frac{1}{2}$ particles}

We now turn  to the more complicate 
case of the system  of two interacting
spin $\frac{1}{2}$ particles 
used in $NMR$ experiments.  Recent 
literature we will refer to on this
topic can be found in \cite{KBG}, \cite{Ramaontheweb}. 
The papers 
\cite{ObWa1}, \cite{ObWa2} introduce 
system-theoretic aspects of 
 $NMR$ spectroscopy.

\vs

The Hamiltonian of a system of 
two interacting spin $\frac{1}{2}$ particles 
which interact with each-other,  
and are immersed in
a driving electro-magnetic field,  is 
given by \cite{BrNa} \cite{venti} \cite{uno} 
\be{Hamiltoniano}
H(t):=\sum_{k=x,y} (\gamma_1I_{1k}+\gamma_2I_{2k})u_k(t)+ 
(\gamma_1I_{1z}+\gamma_2I_{2z})\bar u_z+JI_{1z}I_{2z}. 
\ee
The first term on the right hand 
side of (\ref{Hamiltoniano}) 
represents the interaction of
the two particles with the $x$ and $y$ component of the 
external magnetic field, $u_x$ and $u_y$,  which are allowed 
to vary with time. The second term represents the interaction with
the $z$ component of the field, $\bar u_z$,  which is kept constant in Nuclear
Magnetic Resonance experiments. The last term  represents the interaction
between the two particles which is  modeled with a scalar {\it Ising}
term. 
The constants $\gamma_1$ and $\gamma_2$ are the gyromagnetic
ratios of particle $1$ and $2$,  respectively. 
The constant $J \not=0$ is the coupling constant between the two
particles. For $k=x,y,z$, we have 
\be{ope1}
I_{1k}:= \sigma_k \otimes {\bf 1}, 
\ee
and 
\be{ope2}
I_{2k}:={\bf 1} \otimes \sigma_k, 
\ee
where $\sigma_k$, $k=x,y,z$,  are the 
components of the spin operator in
the $x,y,z$ direction and ${\bf 1}$ 
is the identity operator. Also,  we use the notation  
\be{interaction}
I_{1k}I_{2j}:=\sigma_{k} \otimes \sigma_{j}, \qquad j,k=x,y,z. 
\ee
Schr\"odinger equation for the evolution operator $X$ for this system is
given by\footnote{We are setting the Planck constant $\hbar = 1$.}
\be{scrod}
\dot X=-iH(t)X, 
\ee
where $H(t)$ is given in (\ref{Hamiltoniano}).  
We consider the basis $|++>$, $|+->$, 
$|-+>$, $|- ->$ (spin 1 up, spin 2 up and spin 1 up, spin 2 down, and so on), 
in the underlying four dimensional
Hilbert space. In this basis, the matrix representatives of the tensor 
products in (\ref{ope1}), (\ref{ope2}), (\ref{interaction}) are the 
Kronecker products of the 
$2 \times 2$ matrix representatives of the operators
that appear as factors, where the matrix
representatives of $\sigma_{x,y,z}$ 
are given by the Pauli matrices 
(\ref{Paulix}) (\ref{Pauliy}) (\ref{Pauliz}).  
We can write system (\ref{Hamiltoniano}), (\ref{scrod})  in the form 
\be{formaintermedia}
\dot X=A X+ B_xX u_x(t) + B_yX u_y(t)+ B_z \bar u_zX, 
\ee
with 
\be{bardiA}
A:=-iJI_{1z}I_{2z}, \quad B_{x,y,z}=-i (\gamma_1 S_{x,y,z} \otimes {\bf 1} + 
\gamma_2 {\bf 1} \otimes  S_{x,y,z}),  
\ee
and $\bf 1$ represents the $2 \times 2$ identity matrix. The state 
$X$ of (\ref{formaintermedia}) with initial condition equal to
the identity, varies on the Lie group $SU(4)$. 
It is well known \cite{suss} \cite{Murti} 
that every state in $SU(4)$ can be reached 
from the identity by varying the (arbitrarily bounded) 
control functions $u_x(t)$, $u_y(t)$, 
if and only if the Lie algebra 
generated by $\{A +B_z \bar u_z, B_x, B_y \}$, that we will denote here
by $\cal L$, is equal to $su(4)$. In this case the system is said
to be controllable (see also \cite{confra}, 
\cite{Sol1}, \cite{Sol2} for more explicit controllability 
criteria for quantum systems).   More in
general, the set of states that can be obtained with arbitrarily
bounded controls for system (\ref{formaintermedia}) is the connected
subgroup of $SU(4)$ corresponding to the Lie algebra $\cal L$. 
Two cases can  be considered: the
{\it heteronuclear} ($\gamma_1 \not= \gamma_2$) and the {\it
homonuclear} ($\gamma_1=\gamma_2$). We will prove some general
properties concerning the Lie algebra structure of these systems and in
the next two sections we will study in some detail the controllability
properties and give control algorithms for the system of two homonuclear
spins.  

\vs 

In order to study the structure of 
the Lie Algebra underlying system
(\ref{formaintermedia}), we will 
need the following general result 
on the structure of the Lie Algebra 
$su(n)$, for general $n \geq 2$. The proof 
is based on the  Cartan decomposition \cite{Helgason} 
of the Lie group $SU(n)$ and it is presented in \cite{PPR}. A review of  
the Cartan decomposition for general
semisimple Lie groups can be found in Appendix A. 

\vs 

\noindent{\bf Lemma 4} {\it The subalgebra 
$so(n)$ and every other element 
$f \in su(n)$, 
$f \notin so(n)$,  
generate $su(n)$.}

\vs

\noindent{\bf Proof.} See \cite{PPR}.

\vs 

For an heteronuclear system ($\gamma_1 \not= \gamma_2$), it is easily
seen that, with repeated Lie  brackets of $B_x$ and $B_y$, it is
possible to generate all the elements of the form $-iI_{1k}$,
$-iI_{2k}$ in (\ref{ope1}) (\ref{ope2}).  Therefore,
the Lie algebra generated by $B_x$ and  $B_y$ is given by 
$su(2) \times su(2)$. It is known that $su(2) \times su(2)$ is
isomorphic to $so(4)$, therefore, we would like to use Lemma $4$ to
conclude that, no matter what the matrix modeling the interaction 
$A \notin so(4)$ is, the Lie algebra generated by $A+B_z\bar u_z$, 
$B_x$ and $B_y$ is equal to $su(4)$. This requires a little 
care if no further information is provided. 
Although $su(2) \times su(2)$ is isomorphic to $so(4)$,
it might `sit' in $su(4)$ in a different manner from $so(4)$. It turns
out that this is not the case since $su(2) \times su(2)$ is in fact
conjugate to $so(4)$ via an element of $U(4)$. Therefore we can
conclude with the following result which is independent on how we have
modeled the interaction in the matrix $A$. 

\vs 

\noindent {\bf Theorem 5} {\it For every system of two interacting 
heteronuclear spins the Lie algebra $\cal L$ is equal to $su(n)$ and 
the system  is controllable.}  

\vs 


\vs

For homonuclear spins the Lie algebra $\cal L$ associated to  
the model  
depends on the type of interaction we consider. For a general
interaction of the form $A=-i (aI_{1x}I_{2x}+ bI_{1y}I_{2y}+cI_{1z}I_{2z})$ 
 the Lie algebra  $\cal L$ is isomorphic to $u(2)$, if $a=b=c\not=0$ 
and to
$u(3)$ if at least two of the coefficients $a,b,c$ are different. Let us
consider in detail the case of Ising interaction where $a=b=0$, $c=J
\not=0$.  

The matrices $B_{x,y,z}$ are proportional  to $-i(S_{x,y,z}
\otimes {\bf 1} + {\bf 1} \otimes S_{x,y,z})$ and $B_{x}$ and $B_y$
generate a subalgebra isomorphic to $su(2)$ and $so(3)$. For reasons
that will be clear shortly, we call this subalgebra  $\cal
K$. Now write $A:=-iJI_{1z}I_{2z}$ as 
\be{newA}
A=A_1+ \frac{1}{3}D, 
\ee  
with 
\be{A1ante}
A_1:=-\frac{iJ}{3} (2 I_{1z}I_{2z}-I_{1x}I_{2x}-I_{1y}I_{2y}), 
\ee
and 
\be{D}
D:=-i J\sum_{k=x,y,z} I_{1k}I_{2k}.  
\ee
The matrix $\frac{D}{3}$ commutes with $A_1$ and  $B_{x,y,z}$ (and
therefore with the Lie algebra generated by them). Repeated 
Lie brackets of $A_1$ and  
$B_{x,y,z}$ generate matrices of the form  $-i (S_k \otimes S_j 
+S_j \otimes S_k)$, $j\not=k, j,k \in \{x,y,z\}$  and 
$-i(S_r \otimes S_r-S_m  \otimes S_m)$, $r,m \in \{ x,y,z\}$ that span a
vector space that we denote by $\cal P$. Since $A_1 \in {\cal P}$, $D
\in {\cal L}$.  We define ${\cal G}:={\cal K} \oplus {\cal P}$, and 
we have $[{\cal K},
{\cal K}] \subseteq {\cal K}$, $[{\cal K}, {\cal P}] \subseteq {\cal
P}$, and $[{\cal P}, {\cal P}] \subseteq {\cal K}$. Therefore 
\be{lzj}
{\cal L}=span \{\frac{D}{3} \} \oplus {\cal G},  
\ee  
and the subalgebra ${\cal G}$ has a Cartan decomposition
as described in Appendix $A$.

More information can be obtained on ${\cal L}$  if we perform a change
of coordinates diagonalizing $D$ in (\ref{D})\footnote{$X \rightarrow
TXT^*$ with 
$$
T:=\frac{1}{\sqrt{2}}\pmatrix{0&  i & -i & 0\cr 
0 & 1 & 1 & 0\cr 
-i & 0 & 0 & -i \cr
-1 & 0 & 0 &1}
$$
}. 
This transforms the matrix $D$ into $D=-i \frac{J}{4} diag(-3,1,1,1)$
and all the matrices in ${\cal G}$ into matrices of the
form \be{fjfh}
L:=\pmatrix{0 & 0 \cr 0 & R},
\ee
with $R \in su(3)$. In particular, for  matrices
in $\cal K$, $R \in so(3)$, while for matrices in $\cal P$, 
$R \in {\cal S}$, the vector space of $3 \times 3$, zero trace,
symmetric, purely imaginary matrices. Therefore, ${\cal G}$ 
is isomorphic to $su(3)$ (notice the fact that {\it all of} 
$su(3)$ is generated can be obtained as an application of Lemma 4) 
and $\frac{D}{3}$ plays the role of   $iI_{3 \times 3}$. In summary, 
we have for $\cal L$ 
(the symbol $\approx$ indicates Lie algebra
isomorphism). 
\be{dl}
{\cal L}= span \{\frac{D}{3} \} \oplus {\cal K} \oplus {\cal P}  
\approx u(3) = 
span \{ iI_{3 \times 3} \} \oplus su(3) =span \{ iI_{3 \times 3} \}
\oplus so(3) \oplus {\cal S}.  
\ee

\vs

In the following two sections, we focus on controllability analysis and
control algorithms for the system of two homonuclear spin $\frac{1}{2}$
particles. A number of algorithms 
based on Lie group decompositions can be found in the literature for the
heteronuclear case 
(see e.g. \cite{KBG}, \cite{Rama1}, \cite{Rama2}, 
\cite{Ramaontheweb}).  Starting with a
decomposition in factors of the target 
matrix, some algorithms use the
so called {\it hard pulses} (namely 
very high amplitude controls) 
to obtain approximately 
state transfer within the Lie group corresponding to the Lie algebra
generated by the $B$ matrices in a system like
(\ref{formaintermedia}). This along with zero pulses, which make the
state of the system vary on the one dimensional subgroup corresponding
to the matrix $A$,  can be shown to obtain all the possible targets,  if the
system is underlying a Cartan decomposition. 
While this approach 
has been shown to lead to time optimal
controls \cite{KBG}, it has been criticized because 
of the practical feasibility and possible side  
 effects of the hard pulses. The paper
\cite{Ramaontheweb} contains a detailed 
discussion and a comparison between hard 
pulses and {\it soft pulses}
control as well as  an algorithm for 
the control of the system of two
heteronuclear spin with arbitrarily 
bounded control. We shall present in
Section 5,  an
algorithm that, without any approximation, 
steers the system of two
homonuclear spin $\frac{1}{2}$ 
particles to any configuration. This algorithm 
will be an application of
the decomposition of $SU(2)$ described in Theorem 2.

\section{Controllability analysis} 

Using the Cartan decomposition 
for the Lie group $su(3)$ 
above described, it is possible 
to obtain more information about the
controllability of the system of two homonuclear spins. 
In particular it is
possible to obtain explicit expressions for the reachable sets with
piecewise continuous (but not a priori bounded) control. 
To do so, we will use the
main result of \cite{KBG} 
which is called there the `Time Optimal Tori
Theorem'. In the following, we will denote by ${\cal
R}(t)$ the set of states that can be reached from the identity in time
$t$ and by ${\cal R} (\leq T)$ the set $\cup_{0 \leq t \leq T} {\cal R}
(t)$.  We shall denote by $SU(2)^2$ the Lie
group corresponding to the Lie algebra (isomorphic to $su(2)$ and
$so(3)$) generated by 
$\{ -i(S_k \otimes {\bf 1} + {\bf 1} \otimes S_k), k=(x,y,z)\} $,
namely the Lie group of matrices of the form $L \otimes L$, with $L \in
SU(2)$. This is a not faithful (2 to 1) representation of $SU(2)$ and
therefore isomorphic to $SO(3)$. This can also be seen by a change of
coordinates as pointed out in the previous section.

\vs

\noindent {\bf Theorem 7.} {\it For the system of 
two homonuclear spins, let $D$ be the matrix
defined in (\ref{D}). For every $T > 0$, 
\be{HomoNUC}
clos {\cal R}(T)=e^{\frac{1}{3}DT} \{X_f \in SU(4)|  
X_f=K_1 e^{\alpha_1 A_1+\alpha_2 A_2 + \alpha_3 A_3}K_2 \},
\ee
with $K_{1,2} \in SU(2)^2$,  $\alpha_{1,2,3} \geq 0$,  
$\alpha_1+\alpha_2+\alpha_3=T$,  and 
\be{A1}
A_1:=A-\frac{1}{3}D = -\frac{iJ}{3}(2
I_{1z}I_{2z}-I_{1x}I_{2x}-I_{1y}I_{2y}), 
\ee 
\be{A2}
A_2:=-\frac{iJ}{3}(2 
I_{1y}I_{2y}-I_{1x}I_{2x}-I_{1z}I_{2z}), 
\ee
\be{A3}
A_3:=-\frac{iJ}{3}(2
I_{1x}I_{2x}-I_{1z}I_{2z}-I_{1y}I_{2y}). 
\ee
}

\vs

\noindent {\bf Proof.} See \cite{PPR}.

\vs

The eigenvalues of the matrices $A_1$, $A_2$ and $A_3$ are given by 
\be{eiggg}
\lambda_1=\lambda_2=-\frac{1}{6}Ji, \quad \lambda_3=0, \quad \lambda_4= 
\frac{1}{3}Ji,   
\ee
therefore the functions $e^{A_j t}$, $j=1,2,3$ are periodic with period
$\frac{12 \pi}{|J|}$. This shows that if $T \geq \frac{36 \pi}{|J|}$ in
(\ref{HomoNUC}) $clos {\cal R}(T)=e^{\frac{DT}{3}}{\bf G}$, where $\bf
G$ denotes the Lie group (isomorphic to $SU(3)$) corresponding to the
Lie algebra $\cal G$ isomorphic to $su(3)$ 
generated by $A_1,$ $B_x$, $B_y$ and
$B_z$. With these notations, we have 

\vs

\noindent{\bf Theorem 8.} {\it If $T \geq  \frac{36 \pi}{|J|}$, then 
\be{l1}
{\cal R}(T)=e^{\frac{DT}{3}} {\bf G}
\ee}

\vs 

\noindent {\bf Proof.} See \cite{PPR}. 

\section{Control algorithms for two homonuclear spins}

In this section we give a constructive control algorithm for the system
of two homonuclear interacting spin $\frac{1}{2}$ particles, 
which, in the spirit
of Theorem 3, allows for arbitrary bounds on the control. This algorithm
is based on a Lie group decomposition of the Lie group $SU(3)$, and on
an application of the decomposition result of Theorem 2. .

\vs

In the following we will set  
$\gamma=\gamma_1=\gamma_2$.  Consider 
system (\ref{formaintermedia}) with 
$\gamma_1=\gamma_2$ and define 
\be{U1}
U:=e^{-\frac{1}{3}D t}X. 
\ee
Since $D$ commutes with all the matrices in equation
(\ref{formaintermedia}) we have that $U$ satisfies the equation 
\be{diffforu}
\dot U=A_1U+B_xUu_x+B_yUu_y+B_zU\bar u_z,  
\ee
with $A_1$ given in (\ref{A1}) (sse Theorem 7). The matrices  
$A_1$, $B_x$, $B_y$ and $B_z$ generate a Lie
algebra ${\cal G}$ isomorphic to $su(3)$ 
and in particular they are underlying  a Cartan
decomposition of $ {\cal G}$, in that $A_1 \in {\cal P}$ and  
$B_x$, $B_y$ and $B_z$ generate  $\cal K$, and ${\cal G}= {\cal K}
\oplus {\cal P}$ in the notations of Appendix
$A$. 
It is also
useful to perform a change of coordinates as described in Section 3  
diagonalizing the matrix $D$
in (\ref{D}) as 
\be{newD}
D=-i\frac{J}{4}diag(-3,1,1,1).
\ee 
This transforms $A$
into the matrix $A=-i \frac{J}{4}diag({-1,-1,1,1})$, 
and $B_x$, $B_y$ and
$B_z$ into the matrices 
${\gamma} S_{2,3}$, $-{\gamma} S_{2,4}$ and 
${\gamma} S_{3,4}$, respectively, where $S_{j,k}$, $j<k$,  
denotes the matrix $\in so(4)$ with all zero entries except the $(j,k)$-th
and the $(k,j)$-th that are equal to $1$ and $-1$ respectively. In these
coordinates,  the matrix defined in (\ref{A1}) is given by 
$A_1=-i\frac{J}{4}diag(0,-\frac{4}{3}, \frac{2}{3}, \frac{2}{3})$. 
This shows that the Lie group on which the state of the system 
(\ref{diffforu}) varies is given by matrices of the form 
\be{for}
S=\pmatrix{ 1 & 0 \cr 0 & G}, 
\ee 
with $G \in SU(3)$. Therefore we can assume that 
the target matrix $X_f$
has the form 
\be{targettt}
X_f=e^{\frac{DT_f}{3}}S_f,
\ee
where $D$ is now in the new coordinates, 
$T_f\geq 0$ and $S_f$ has the form  in
(\ref{for}). 
\vs
In the system of differential equations in the new coordinates 
\be{inthenew}
\dot X=AX+B_xu_xX+B_yu_yX+B_z\bar u_zX, 
\ee
it will be convenient to 
 scale the time $t$ by a factor $\frac{J}{6}$
and redefine the control 
functions as $u_x \rightarrow \gamma
\frac{6}{J}u_x (\frac{6}{J}t)$, $u_y \rightarrow -\gamma
\frac{6}{J}u_y (\frac{6}{J}t)$, $\bar u_z \rightarrow  \gamma
\frac{6}{J} \bar u_z$, which gives simpler expressions for the matrices $A$, 
$B_{x,y,z}$. In particular, we have that the matrix $B_{x,y,z}$ become 
\be{BXBY}
B_{x,y,z} =\pmatrix{0 & 0 \cr 
0 & S_{({1,2}),({1,3}), ({2,3})}} 
\ee 
where $S_{j,k}$ are the standard basis matrices in $so(3)$. The matrix $D$
becomes $D=-i \frac{3}{2} diag(-3,1,1,1)$ and the 
matrix $e^{Dt}$ is periodic with period 
$\frac{4 \pi}{3}$  and therefore $T_f$ in (\ref{targettt}) 
is defined up to multiples of $4\pi$. The matrix $A_1$ takes the simple
form $A_1:=A-\frac{D}{3}=i diag (0,2,-1,-1)$. We shall refer to this
time scale and control variables in the rest of our treatment.  

\vs 

We now assume that $T_f$ and $S_f$ are given and we show how to steer 
the state of the system to the desired final configuration
(\ref{targettt}) with
arbitrary bound on the control. We 
first remark that the problem amounts
to the following one: 

\vs

\noindent{\bf Problem 1} {\it Steer the state of the system 
\be{igh}
\dot S=A_1S+B_xSu_x+B_ySu_y,
\ee
from the identity to 
$e^{-B_z \bar u_z (T_f+n4\pi)}S_f$ in time $T_f+4n\pi$ for
sufficiently large $n$, using piecewise 
constant controls of the form 
$u_x \equiv 0$, $u_y \equiv const$, or 
$u_x \equiv const$, $u_y \equiv 0$.}

\vs 
To see this, define 
\be{esse}
S:=e^{(-\frac{1}{3}D-B_z \bar u_z)t} X, 
\ee
and using (\ref{inthenew}), $S$ satisfies 
\be{peresse}
\dot S=A_1S+e^{-B_z \bar u_z t} B_y e^{B_z \bar u_z
t}Su_y(t)+
e^{-B_z \bar u_z t} B_x e^{B_z \bar u_z
t}Su_x(t). 
\ee
In deriving (\ref{peresse}) we have used the fact that $B_z$ commutes
with $A$ and $D$ commutes which each one of the matrices in equation
(\ref{inthenew}). Now a direct calculation shows that if
$u_x(t)=\bar k cos(\bar u_z t)$ and 
$u_y=-\bar k sin (\bar u_z t)$ then the last  two elements in the above
equation give $\bar k B_xS$, and if $u_x(t)=\bar k cos(\bar u_z t)$ and 
$u_y(t)=\bar k sin(\bar u_z t)$ the last two terms give 
$ \bar k B_yS$. Therefore, if
one keeps switching between these two types of controls the right hand
side of equation (\ref{peresse}) is alternatively of the form 
$A_1S+\bar k B_x S$ and   $A_1S+\bar k B_y S$, where we have denoted by
$\bar k$ any constant. Therefore the choice of the control reduces to
the choices of the constants $\bar k$. These calculations could have
been carried out in the original coordinates and are often referred to
in the physics literature as `considering the system in a rotating frame'
\cite{Ramaontheweb}. We emphasize  that the calculations in the case
considered here do not involve any approximation.

Now, if we are able to steer the 
state of the system (\ref{igh}) or
equivalently the state of (\ref{peresse}) 
with the above described controls to $e^{-B_z \bar u_z (T_f +4n\pi)} S_f$ 
in time $T_f+4n\pi$, it follows from  (\ref{esse}) that the state of the
original system (\ref{inthenew}) is driven to $e^{\frac{1}{3}DT_f+ 
\frac{4nD\pi}{3}} S_f=e^{\frac{DT_f}{3}} S_f$, because of the
periodicity of $e^{Dt}$.

\vs

{\bf Solution to  Problem 1} 

\vs 

Notice that, in the new coordinates, the first rows and columns of
the matrices $A_1$, $B_x$ and $B_y$ are zeros, so that we can consider
the system as a $3 \times 3$ one with state varying on $SU(3)$. 

\vs

We show next, how to obtain every matrix of the form
$e^{B_xt}=e^{S_{1,2}t}$ (notations are for $3 \times 
3$ matrices). Set $u_y\equiv 0$ and write $A_1=-F+\tilde A_1$, with
$F=-i diag(\frac{1}{2}, \frac{1}{2},-1)$ and 
$\tilde A_1=-i diag (\frac{-3}{2}, \frac{3}{2},0)$. Setting 
\be{tildeS}
\tilde S:=e^{Ft}S
\ee 
and noticing that $F$ commutes with both $A_1$ and
$B_x$, we obtain 
\be{AHH}
\tilde S=\tilde A_1 \tilde S+B_x \tilde S u_x.  
\ee       
Now, since the third rows and columns of system (\ref{AHH}) are all
zeros the system is essentially 
$2 \times 2$ and the matrices are
essentially in $su(2)$. Therefore, we 
can apply Theorems $2$ and $3$  to
steer the state of this system 
from the identity to the matrix\footnote{An alternative here could have
been  to apply the algorithm of \cite{Rama1} appropriately modified to
allow bounds on magnitude rather than power. This is possible since in
this case the two matrices $\tilde A_1$ and $B_x$ are orthogonal.}  
\be{EBX}
e^{B_x t}:=\pmatrix{cos(t) & sin(t) & 0 \cr 
-sin(t) & cos(t) & 0\cr
0 & 0 & 1}.  
\ee
Let $T_x=T_x(t)$, the time needed 
to steer to $e^{B_xt}$ with bound on
the control,  according to 
Theorems 2 and 3. Now, notice that 
$e^{-B_x \frac{\pi}{2}} \tilde 
A_1 e^{B_x \frac{\pi}{2}}= -\tilde A_1$ and therefore, for every $\tau
\geq 0$, we have 
\be{aspettando}
e^{\tilde A_1 \tau} e^{-B_x \frac{\pi}{2}}e^{\tilde A_1 \tau}
e^{B_x \frac{\pi}{2}}= I_{3\times 3}. 
\ee   
Therefore, we can steer to the identity in time 
\be{TIDE}
T_{Id}(\tau):=2 \tau + T_x(\frac{\pi}{2}) +T_x(\frac{3 \pi}{2})
\ee 
with $\tau \geq 0$
arbitrary. We would like to steer the state of (\ref{AHH}) 
to $e^{B_xt}$ in time $4\bar n\pi$, for some integer $\bar n$. 
We can do that by steering in time $T_x(t)$ 
to $e^{B_xt}$ and then by steering 
(from the identity) to the identity in time $T_{Id}(\tau)$ 
in (\ref{TIDE}), choosing $\tau$ so that
\be{pippo}
T_x(t)+T_{Id}(\tau)=4 \bar n\pi.
\ee 
Notice that, from the decomposition of Theorem 2, the time $T_x(t)$ is
uniformly bounded (as $0 \leq t \leq 2 \pi$) by a value  
$\bar T_x$ that depends on
the bound on the control.  $\bar T_x$ can be easily 
calculated by knowing the order of generation in terms of the two
generating one-dimensional subgroups and knowing their periods. 
Therefore, we can
always assume that $\bar n$ is chosen so that we can make (\ref{pippo})
satisfied for every $t$, $ 0 \leq t \leq 2 \pi$, for appropriate $\tau$,
namely $4 \bar n \pi \geq \bar T_x+T_x(\frac{\pi}{2})+ T_x(\frac{3
\pi}{2})$. 
We will assume this to be the case in the following. 

Now,  since 
$e^{Ft}$ is periodic with period $4 \pi$, by 
steering the state $\tilde S$ of (\ref{AHH}) 
to $e^{B_xt}$ in time $4 \bar n\pi$, using (\ref{tildeS}),  
we have steered the state of the system (\ref{igh}) to $e^{B_xt}$ in
time $4 \bar n \pi$.  
 
\vs 

In a completely analogous way,  one can obtain matrices of the form 
$e^{B_yt}=e^{S_{1,3}t}$ in time $4 \bar n \pi$ (in fact the procedure is
completely equivalent and the value of $\bar n$ is the same in the two 
cases). Moreover, noticing that 
\be{BXP2}
e^{B_x \frac{\pi}{2}} e^{B_yt} e^{-B_x \frac{\pi}{2}}=e^{B_zt}:=e^{S_{2,3}t},
\ee
one can obtain every matrix of the form $e^{B_zt}$ in time $3 \times
4\bar n\pi$. 

\vs 

At this point, we recall a decomposition of the matrices $U_f$ 
$\in SU(3)$ presented in \cite{MURNA}. Every matrix $U_f \in SU(3)$ can
be written as 
\be{Uffa}
U_f=D(\alpha_1,\alpha_2,\alpha_3)U_{12}(\theta_1,\sigma_1) 
U_{13}
(\theta_2, \sigma_2) U_{23}(\theta_3, \sigma_3),
\ee
where the 
matrices $U_{kl}(\theta,\sigma)$, with $k <l$ are the matrices
whose submatrix intersection of the $k-$th and $l-$th rows and columns
is given by 
\be{elle}
L:=\pmatrix{cos(\theta) & -sin(\theta)e^{-i \sigma} \cr 
sin(\theta) e^{ i \sigma } & cos(\theta)}, 
\ee
so that, for example 
\be{U12thetasigma}
U_{12}(\theta_1, \sigma_1):=\pmatrix{cos(\theta_1) & 
-sin(\theta_1) e^{-i\sigma_1} & 0\cr
sin(\theta_1)e^{i\sigma_1} & cos(\theta_1) & 0 \cr
0 & 0 & 1}. 
\ee
The matrix $D(\alpha_1,\alpha_2,\alpha_3)$ is of the form 
\be{MATD}
D(\alpha_1,\alpha_2,\alpha_3) := \pmatrix{ e^{i\alpha_1} & 0 & 0 \cr 
0 & e^{i\alpha_2}& 0 \cr
0 & 0 & e^{i \alpha_3}}, 
\ee
with $\alpha_1 +\alpha_2 +\alpha_3=0$. Once $U_f$ is given, the
parameters $\theta_j,\sigma_j,\alpha_j$, $j=1,2,3$, can be easily
calculated by an analytic procedure described in
\cite{MURNA}. Therefore, specifying $U_f$ is 
equivalent to specifying
the parameters $\theta_j, \alpha_j, \sigma_j$. 

\vs

Now notice that 
\be{llkp}
U_{12}(\theta_1, \sigma_1) = e^{A_1 \frac{-\sigma_1}{3}} 
e^{B_x \theta_1}e^{A_1 \frac{\sigma_1}{3}}.  
\ee
Therefore the  matrix $ U_{12}(\theta_1, \sigma_1)$ can be obtained
(assume w.l.g. $\sigma_1 \geq 0$) by letting the system go with $u_x
\equiv u_y\equiv 0 $, 
for time $\frac{\sigma_1}{3}$, and then drive the system to 
$e^{B_x \theta_1}$ as explained above, and then let the system go again
for time $2 \pi - \frac{\sigma_1}{3}$, setting $u_x \equiv u_y \equiv 0$
again. 
The same thing can be done for the terms $U_{13}
(\theta_2, \sigma_2)$ and $U_{23}(\theta_3, \sigma_3),$ in 
(\ref{Uffa}), with $e^{B_y \theta_2}$ and 
$e^{B_z \theta_3}$ replacing $e^{B_x \theta_1}$, respectively. Therefore
the last three factors in (\ref{Uffa}) can be written as
\be{Mikos}
D(-\alpha_1,-\alpha_2, -\alpha_3)U_f=e^{A_1 t_1} 
e^{B_x t_2} e^{A_1 t_3} e^{B_yt_4} e^{A_1 t_5} e^{B_z t_6} 
e^{A_1 t_7}.  
\ee
Our previous discussion shows how to calculate the parameters $t_j \geq
0$, $j=1,...,7$ and how to drive the identity to each one of the factors
in (\ref{Mikos}). Recalling 
periodicity of the matrix $e^{A_1t}$ and the times to drive to the
matrices $e^{B_xt_2}$, $e^{B_yt_4}$, $e^{B_zt_6}$, calculated in the
above discussion, the transfer to a matrix of the form 
(\ref{Mikos}) takes {\it at most} time 
\be{Tbat}
\bar T_1=4 \times 2 \pi + 2 \times 4 \bar n\pi+ 3 \times 4 \bar n\pi, 
\ee 
where the three terms on the right hand side refer 
to the matrices of 
the form $e^{At}$, $e^{B_xt}$ and 
$e^{B_yt}$, and $e^{B_zt}$,
respectively. Now, the matrix $D(\alpha_1, 
\alpha_2, \alpha_3)$, with
$\alpha_1+\alpha_2+\alpha_3 =0$, 
can be obtained as  
\be{DA}
D(\alpha_1, \alpha_2, \alpha_3)= 
e^{B_x \frac{\pi}{2}} e^{A_1 t_1} 
e^{-B_x \frac{\pi}{2}} e^{A_1 t_2}, 
\ee
if $2t_2-t_1=\alpha_1$ and 
$2t_1-t_2=\alpha_2$. Therefore the 
overall transfer of the state of 
(\ref{igh}) from the 
identity to $U_f$ takes {\it at most} time 
\be{TIME}
\tilde T=\bar T_1+ 2 \times 4\bar n\pi + 2 \times  2 \pi,   
\ee
with $\bar T_1$ defined in (\ref{Tbat}). 

\vs

\noindent Now notice that, for every $\tau \geq 0$, we have 
\be{klm}
e^{B_y \frac{\pi}{2}} e^{A_1 \tau} 
e^{-B_y \frac{\pi}{2}} e^{B_x \frac{\pi}{2}} 
e^{A_1 \tau} e^{-B_x \frac{\pi}{2}} e^{A_1 \tau} = I_{3 \times 3}. 
\ee
Therefore the Identity matrix can be obtained in time 
$3\tau +4 \times 4 \bar n \pi$, for arbitrary $\tau \geq 0$. 

\vs 

Now choose $n$ so that 
\be{lll}
T_f+4n\pi \geq \tilde T+ 4 \times 4 \bar n \pi, 
\ee
and express the matrix $e^{-B_z \bar u_z(T_f+4 n\pi)} S_f$ as $U_f$ in
(\ref{Uffa}). Then, we have showed how  in time at most $\tilde T$ we can 
steer to this final condition. Let $\hat T$ 
($\leq \tilde T$) the time actually used to steer to 
$e^{-B_z \bar u_z(T_f+4 n\pi)} S_f$, and let us define 
\be{kol}
\bar t :=T_f + 4n \pi - \hat T - 4 \times 4 \bar n \pi \geq 0,\ee 
from (\ref{lll}). Then by steering to the identity in time 
$\bar t+ 4 \times 4 \bar n \pi$, we have driven the state of the system
(\ref{igh}) to $e^{-B_z \bar u_z(T_f+4 n\pi)} S_f$, in time 
 exactly $T_f+4 n\pi$,  as desired. \qed

\section{Conclusions}

We have presented control algorithms and controllability analysis for
the systems of one and two homonuclear spin $\frac{1}{2}$ particles in a
magnetic fields. In particular explicit expressions for the reachable
sets have been obtained. 

The control algorithms are based on a Lie group decomposition of the Lie
group $SU(2)$ for which we have obtained explicit expressions of the
parameters involved. All the algorithms described allow for arbitrary
bounds on the controls and provide explicit expressions for the
parameters involved in the control design. Moreover, the algorithms
allow some flexibility in the final time that can be fixed a priori as
long as it is greater than a given value. This is due to the
right invariance of the systems considered and to the fact that the
identity matrix can be obtained in sufficiently large but otherwise
arbitrary 
time. Questions of optimization of the time transfer 
have not been considered here and will be object of future
research. Also, the explicit generation of control protocols which use
the decompositions of $SU(2)$ described here as a basic building block
for different and more complicated systems will 
be investigated.

\section*{Appendix A: Review of Cartan decomposition of Lie groups}

We review, in this Appendix some basic facts about decompositions 
of Lie groups based on symmetric spaces, also known as Cartan
decomposition, that we will use 
in the following. We refer to the texts 
 \cite{Helgason} \cite{Hermann} for further details and generalizations 
as well as for some of the terminology that we use here. Applications to
control theory are considered in \cite{CSL1} \cite{CSL2}.

Consider a semisimple Lie algebra 
$\cal G$ and the corresponding
connected Lie group ${\bf G}$. Assume that 
$\cal G$ admits a decomposition as
a vector space 
\be{decom}
{\cal G}={\cal K} \oplus {\cal P}, 
\ee
where ${\cal K}$ is subalgebra of ${\cal G}$, namely 
$[{\cal K},{\cal K}] \subseteq  {\cal K}$. 
Moreover, assume that the following commutation relations hold 
among the elements of ${\cal K}$ and ${\cal P}$, 
\be{commu1}
[{\cal K},{\cal P}] \subseteq  {\cal P}, 
\ee
\be{commu2}
[{\cal P}, {\cal P}] \subseteq  {\cal K}.  
\ee
Denote by $\bf K$ the connected Lie group 
corresponding to $\cal K$ and assume it is compact. 
Denote  by $\bf P$ the 
image of $\cal P$ under  the exponential map.     
Under the above assumptions and definitions, 
\be{deco1}
{\bf G}={\bf PK}, 
\ee
namely every element of $\bf G$ can 
be written as the product of an element
of $\bf P$ and an element of $\bf K$. 

\vs

Every element in ${\cal P}$ 
belongs to a Cartan subalgebra (namely a 
maximal Abelian subalgebra) $\cal A$ \cite{CSL1} whose dimension is called 
the {\it rank} of the Lie group $\bf G$. Moreover, any two Cartan subalgebras
are conjugate via elements in $\bf K$, namely, if $\cal A'$ is another 
Cartan subalgebra, then there exists an element $K \in \bf K$ such that 
$ KAK^{-1} \in {\cal A}'$, 
for each $A\in {\cal A}$. As a consequence, 
every element $P$ in $\bf P$, can be written as 
$P=K A  K^{-1}$, where $K$ is an element of 
$\bf K$ and $A$ is an element of $\bf A$, 
the connected Lie subgroup of $\bf G$
 associated to $\cal A$. Therefore,  one can write 
\be{deco2}
{\bf G}={\bf KAK}, 
\ee  
namely, every element in $\bf G$ can be written as the product of an element
in $\bf K$, an element in $\bf A$ and an element in $\bf K$, in that order. 

\vs

The simplest example of Cartan decomposition 
(\ref{deco2}) is Euler decomposition for $SU(2)$.
Consider the Pauli matrices in (\ref{Paulix}), (\ref{Pauliy}), 
(\ref{Pauliz}), divided by $i$, $-iS_x$, $-iS_y$, $-iS_z$. These
matrices satisfy the commutation relations  (\ref{commuPauli}).  
Therefore, one can take, for example,  
  ${\cal K}:= span \{S_y\}$, and ${\cal P}:= span \{ S_x, S_z \}$, and 
${\cal A}:=S_z$, although every other combination would be possible. 

\vs 

The decomposition (\ref{deco2}) can be continued by decomposing $\bf K$
in the same fashion as $\bf G$. Continuing this way, one end
up with an expression of every element of the Lie group $\bf G$ in terms
of the product of elements belonging to one dimensional subgroups.

\section*{Appendix B: A decomposition of $SO(3)$}

In the following we shall call $S_{hk}$, $h <k$  
the matrix $\in so(3)$ which has
zeros everywhere except in the $h,k$-th ($k,h$-th) 
entry which is  equal to $1$ ($-1$). 
Given a matrix $Z_1$, there exists a matrix $T_1 \in SO(3)$ such that 
\be{trans1}
T_1Z_1T_1^T:=\lambda_1 S_{12},   
\ee
$\lambda_1>0$. 
This can be easily seen by choosing $T_1:=[v_1, v_2,v_3]^T$, with $v_3$
such that $v_3^T Z_1=0$ and with norm equal to one and $v_1$ and $v_2$ 
such that $\{ v_1,v_2,v_3 \}$ form an orthonormal basis 
in $\RR^3$. We also set 
\be{setting}
T_1Z_2T_1^T:=a S_{12}+ b S_{13}+c S_{23}. 
\ee 
Choose now $T_2:=e^{S_{12} \theta}$ with $\theta$ such that 
$b cos(\theta)+c sin(\theta)=0$.  Then we have 
\be{another}
T_2 T_1 Z_1 T_1^T T_2^T:= \lambda_1 S_{12}, 
\ee 
\be{EA1}
T_2 T_1 Z_2 T_1^T T_2^T:=a S_{12} + d S_{23}, 
\ee
for some parameter $d \not= 0 $. Therefore, we can always assume that,
in appropriate coordinates, the matrices $Z_1$ and $Z_2$ have the form 
$Z_1:= \lambda_1 S_{12}$ and 
$Z_2:= aS_{12} +d S_{23}$, respectively. Moreover we
can divide $Z_1$ by $\lambda_1>0$ (this has the only effect that, in the
matrices if the form $e^{Z_1 t}$,  $t$ has to be scaled by a factor
$\lambda_1$) and analogously we can divide $Z_2$ 
(in the new coordinates in (\ref{EA1})) by $d$ and therefore the
parameter $t$ in the subgroup $e^{Z_2t}$ has to be scaled by a factor
$d$. Define $\rho:=\frac{a}{d}$, we can assume, without loss of
generality, that the matrices $Z_1$ and $Z_2$ are given by 
$Z_1:=S_{12}$ and $Z_2:=\rho S_{12} + S_{23}$, and we shall do so in
the following. Notice also that the above manipulations do not modify the
value of the parameter $\psi$ in (\ref{psi}) which is given, in terms of
$\rho$, by 
\be{EA2}
\psi:=\frac{\rho}{\sqrt{1+ \rho^2}}. 
\ee

\vs

To express a matrix $X_f \in SO(3)$ as 
\be{EA3}
X_f=e^{Z_1 t_1} e^{Z_2 t_2} e^{Z_1 t_3} \cdot  \cdot \cdot e^{Z_1 t_s}, 
\ee
we first recall (see e.g. \cite{sakurai}) 
 that we can express every matrix $X_f \in SO(3)$ as 
\be{EA4}
X_f=e^{S_{12} \alpha} e^{S_{23} \beta} e^{S_{12} \gamma}, 
\ee
with $\alpha, \gamma \in [0,2\pi]$ and $\beta \in [0,{\pi}]$. 
This is the
classical Euler decomposition of a rotation. The Euler 
parameters $\alpha,$ $\beta$ and $\gamma$ can be easily calculated by
using an analytic procedure described  in \cite{nonloso}. Now,
since we have set $Z_1=S_{12}$, the problem is to express every matrix
of the form $X_f=e^{S_{23} \beta}$ as in (\ref{EA3}). To do that, we
show how to express every matrix of the form 
$e^{S_{23} \frac{\beta}{m}}$ as 
\be{EA5}
e^{S_{23} \frac{\beta}{m}}=e^{Z_1 t_1} e^{Z_2 t_2} e^{Z_1 t_3}, 
\ee 
for sufficiently large $m$. An explicit calculation gives 
\be{EA6}
e^{Z_2 t_2}=\pmatrix{\frac{1+\rho^2 c}{\eta^2} & \frac{\rho s}{\eta} & 
\frac{\rho -\rho c}{\eta^2} \cr
\frac{-s \rho}{\eta} & c & \frac{s}{\eta} \cr
\frac{-c \rho + \rho}{\eta^2} & \frac{-s}{\eta} & \frac{c+
\rho^2}{\eta^2}},   
\ee
where $\eta:=\sqrt{1+\rho^2}$, $s:=sin(\eta t_2)$, $c:=cos(\eta
t_2)$. Now, if we choose $m$ so that 
\be{EA7}
2 \psi^2 -1 \leq cos(\frac{\beta}{m}), 
\ee
we can choose $t_2$ so that 
\be{EA8}
\frac{c+ \rho^2}{\eta^2}:=\frac{cos(\sqrt{1+ \rho^2} t_2) + \rho^2}{1+
\rho^2} =cos(\frac{\beta}{m}). 
\ee
Assume that this is the case and consider $e^{Z_2 t_2}$ in (\ref{EA6})
with this choice. For brevity, let us call $a_{ij}$ its $i,j$-th
element. By choosing $t_1$ so that $cos (t_1) a_{13}+ sin (t_1)
a_{23}=0$, we can
make the $1,3$-th entry of the matrix $e^{Z_1t_1}e^{Z_2t_2}$ equal to
zero. The $2,3-$ entry of this matrix is  equal to 
$sin(t_1) a_{13} + cos(t_1) a_{23}$ and it is equal to $sin(\frac{\beta}{m})$ up to the sign that
can be changed by replacing $t_1$ with $t_1 +\pi$. 
Let us now denote by $a_{ij}$ again the entries
of the new matrix  $e^{Z_1t_1}e^{Z_2t_2}$. By choosing $t_3$ so that 
$a_{11} sin(t_3)+a_{12}cos(t_3)=0$, we can make the $1,2$-th entry of the
matrix   $e^{Z_1t_1}e^{Z_2t_2}e^{Z_1 t_3}$ equal to zero, without
affecting the $1,3-$th, $2,3-$th and $3,3$-th entries 
(this procedure is reminiscent of the
standard calculation of Euler angles in \cite{nonloso}). The resulting
matrix is exactly $e^{S_{23} \frac{\beta}{m}}$. Therefore, by considering $m$ factors, 
a constructive decomposition of $SO(3)$ has been obtained.

\end{document}